\documentclass[aps,prb,reprint,groupedaddress,amsmath,amssymb,showpacs]{revtex4-1}

\usepackage[colorlinks=true,allcolors=blue,breaklinks]{hyperref}
\usepackage{graphicx}
\usepackage{dcolumn}
\usepackage{bm}
\usepackage{nicefrac}
\usepackage{stackengine}
\usepackage{amsmath}

\stackMath

\begin{document}

\title{Phase transition dynamics in the three-dimensional field-free $\pm${\textit J} Ising model}
\author{Ozan S. Sar{\i}yer}
\email[]{ossariyer@pirireis.edu.tr}
\affiliation{P\^{\i}r\^{\i} Reis University, School of Arts and Sciences, Istanbul 34940 Turkey}
\date{\today}

\begin{abstract}
By using frustration-preserving hard-spin mean-field theory, we investigated the phase transition dynamics in the three-dimensional field-free $\pm J$ Ising spin glass model. As the temperature $T$ is decreased from paramagnetic phase at high temperatures, with a rate $\omega=-dT/dt$ in time $t$, the critical temperature depends on the cooling rate through a clear power-law $\omega^a$. With increasing antiferromagnetic bond fraction $p$, the exponent $a$ increases  for the transition into the ferromagnetic case for $p<p_\text{c}$, and decreases for the transition into the spin glass phase for $p>p_\text{c}$, signaling the ferromagnetic-spin glass phase transition at $p_\text{c}\approx0.22$. The relaxation time is also investigated, at adiabatic case $\omega=0$, and it is found that the dynamic exponent $z\nu$ increases with increasing $p$.
\end{abstract}


\maketitle

\section{Introduction}

Spin glass systems exhibit aging, rejuvenation and memory effects owing to their long relaxation times. \cite{Dupuis01} Better understanding of the spin dynamics in such systems is a key for applications such as information storage, magnetic resonance imaging, biomarking, biosensing, and targeted drug delivery such as hyperthermia in cancer treatment. \cite{Perovic13} Materials such as Fe$_2$TiO$_5$, YEr (2\%) and Fe$_{0.5}$Mn$_{0.5}$TiO$_3$ are classified as \emph{Ising spin glasses}. \cite{Svedlindh87a}

In this paper, we study the dynamic effects on phase transitions in 
the $\pm J$ random-bond Ising spin glass model in zero external magnetic field that is defined on a lattice by the dimensionless Hamiltonian \cite{Toulouse77}
\begin{equation}
\label{eq:Model}
\mathcal{H}=-\sum_{\langle ij\rangle}J_{ij}s_is_j.
\end{equation}
Here, the sum is over $ij$-bonds, $s_i=\left\{+1,-1\right\}$ is the Ising spin at $i$-site, and $J_{ij}=\left\{+J,-J\right\}$ is the $ij$-bond interaction dealt quench-randomly, with an antiferromagnetic ($-J$) probability $p$. A dimensionless temperature for the system can be defined as $T=1/J$.

The hard-spin mean-field theory (HSMFT) is defined by the set of self-consistent equations
\begin{equation}
\label{eq:Method}
m_i=\sum_{\left\{s_j\right\}}\left\{\left[\prod_j\frac{1+m_js_j}{2}\right] ~ \tanh\left(\sum_jJ_{ij}s_j\right)\right\}
\end{equation}
for the local magnetization $m_i$ at each $i$-site, with nearest-neighbors labeled by $j$. For it preserves the local frustration, it is useful in studying spin glass systems. \cite{Banavar91, Netz91a, Netz91b, Netz91c, Netz92, Netz93, Ames94, Berker94, Kabakcioglu94, Akguc95, Tesiero96, Monroe97, Pelizzola99, Kabakcioglu00, Kaya00, Yucesoy07, Robinson11, Caglar11, Caglar15, Sariyer12}

\begin{figure}[t!]
\includegraphics[scale=1.0]{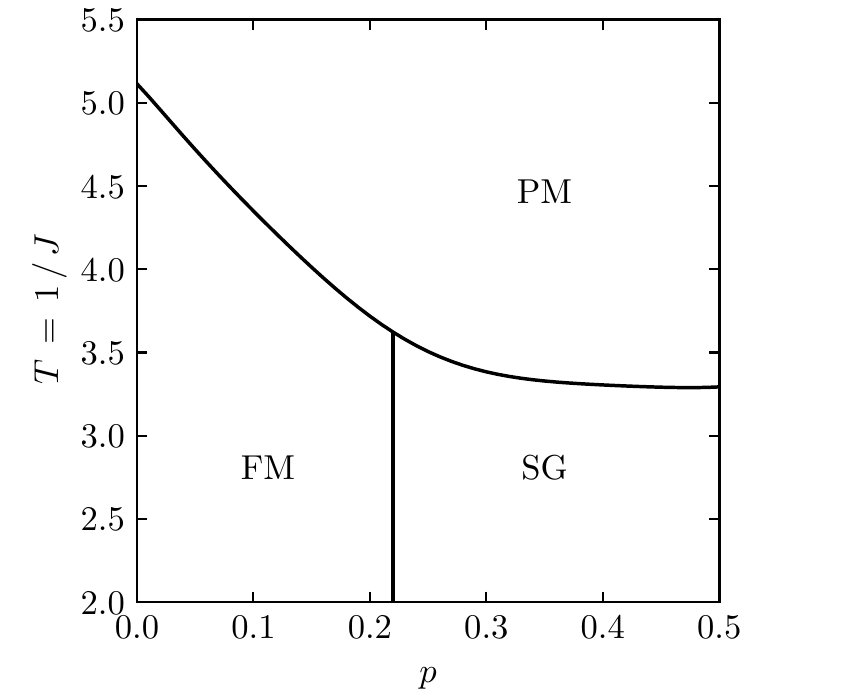}
\caption{\label{fig:PhaseDiag} Zero-field phase diagram of the $\pm J$ model on a simple cubic lattice. PM, FM, and SG respectively denote paramagnetic, ferromagnetic, and spin glass phases.}
\end{figure}

Previously, we studied the $\pm J$ model on a simple cubic lattice under time-dependent external magnetic field by HSMFT, and obtained the phase diagram (see Fig.~\ref{fig:PhaseDiag}) from distinctive behavior of hysteresis loops. \cite{Sariyer12} At temperatures above a critical $T_\text{c}^0$ that depends on $p$, system is in paramagnetic (PM) phase. (In general, the critical temperature $T_\text{c}(\omega)$ depends on the cooling rate $\omega$, and $T_\text{c}^0\equiv T_\text{c}(\omega=0)$ denotes the non-dynamic critical temperature at infinitely-slow cooling rate.) At low temperatures, $T<T_\text{c}^0(p)$, there exist a ferromagnetic (FM) phase for $p<p_\text{c}$, and a spin glass (SG) phase for $p>p_\text{c}$. The FM-SG phase boundary is at $p_\text{c}\approx0.22$. The phase diagram of Fig.~\ref{fig:PhaseDiag} is consistent both with theoretical \cite{Ozeki87} and experimental \cite{Binder86} results, and is mirror symmetric about $p=0.5$, with antiferromagnetic (AFM) phase replacing the FM phase for $p>1-p_\text{c}$. We can omit the $p>0.5$ subspace in general, due to the FM$\leftrightarrow$AFM symmetry of classical Ising spins on bipartite lattices.

In the same previous work, dynamic effect of the magnetic field sweep rate on the dissipative loss in a hysteresis cycle was also studied \cite{Sariyer12}, while in the present article, we consider the dynamics of the model by means of (\emph{i}) adiabatic relaxation time behavior for infinitely-slow cooling rate $\omega=0$, and (\emph{ii}) the effects of nonzero cooling rate $\omega$ on the critical temperature $T_\text{c}(\omega)$.

To this end, we consider an $L\times L\times L$-site simple cubic lattice with periodic boundary conditions. HSMFT results become independent of the lattice size for $L>15$ for this system \cite{Yucesoy07, Sariyer12}, and we work with $L=20$ here.

\section{Relaxation time}

A particular realization at a given $(T,p)$ is generated by the assignment of the quenched-random interactions $J_{ij}=\pm1/T$ with a probability $p$ of AFM bonds $-1/T$, and initially, an unbiased random assignment of spins in the range $s_i\in\left[-1,+1\right]$. A time step $dt$ corresponds to successive iterations of Eq.~(\ref{eq:Method}) on $L^3$ randomly chosen sites. The relaxation time $\tau$ corresponds to number of such time steps required for the system to converge from an initial random seed to a self-consistent solution that is determined when the mean-square deviation in magnetization, $\Delta m^2=L^{-3}\sum_i\left[m_i(\tau)-m_i(\tau-1)\right]^2$, drops below a predetermined threshold, \emph{i.e.}, when $\Delta m^2\leq10^{-12}$.

\begin{figure}[t!]
\includegraphics[scale=1.0]{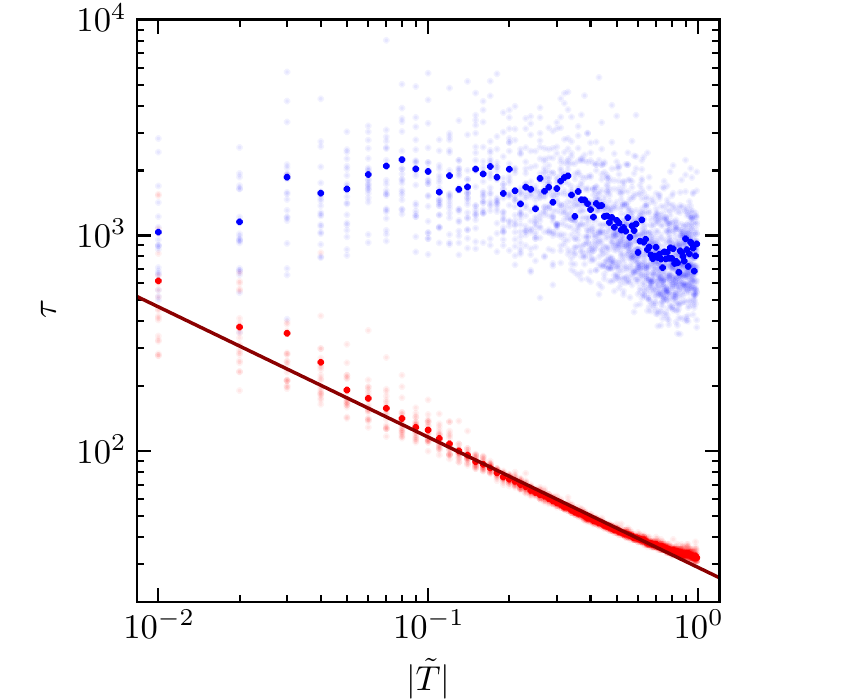}
\caption{\label{fig:RelaxTime} Relaxation time $\tau$ as a function of absolute reduced temperature $|\tilde{T}|=|T-T_\text{c}^0|/T_\text{c}^0$ for $p=0.25$. Faded markers show the results for $20$ distinct realizations of AFM bond distributions, while the dark markers show averages over those random replicas. Blue and red show the data below and above the critical temperature, $\tilde{T}<0$ and $\tilde{T}>0$ respectively. Line is the best fit to the power-law Eq.~(\ref{eq:RelaxTime}), for $\tilde{T}\geq0.1$.}
\end{figure}

\begin{figure}[t!]
\includegraphics[scale=1.0]{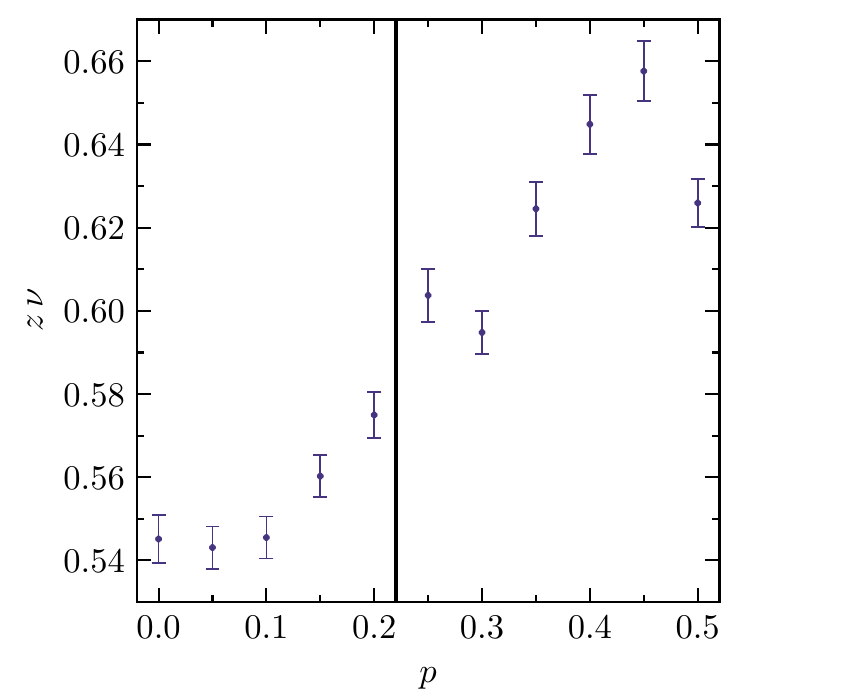}
\caption{\label{fig:DynamicExp} Dynamic critical exponent $z\nu$ as a function of AFM bond fraction $p$. Error bars show the standard deviation of the fit procedure to Eq.~(\ref{eq:RelaxTime}). Vertical line shows the critical $p_\text{c}\approx0.22$ between the ordered FM and SG phases (see Fig.~\ref{fig:PhaseDiag}).}
\end{figure}

A characteristic signature of spin glass systems is that the relaxation time $\tau$ diverges as $T\to{T_\text{c}^0}^+$, and remains infinite for $T<T_\text{c}^0$. Fig.~\ref{fig:RelaxTime} is a log-log plot of the relaxation time $\tau$ as a function of $|\tilde{T}|$, for a representative value $p=0.25$, while we obtain the same behavior for other $p$ values. Here,
\begin{equation}
\tilde{T}\equiv\frac{T-T_\text{c}^0}{T_\text{c}^0}
\end{equation}
is the reduced temperature that measures the distantness to critical temperature. Faded markers show the results for $20$ distinct realizations of AFM bond distributions, while the dark markers show averages over such random realizations. We see that the relaxation time $\tau$ averaged over replicas, peaks as $\tilde{T}\to0^+$ and stays at large values for $\tilde{T}<0$ as expected. Deviation from a true divergence at $\tilde{T}\leq0$ is due to finite-size effects.

The divergence of $\tau$ as $\tilde{T}\to0^+$, referred as the critical slowing-down, is due to the divergence of the correlation length $\xi$. A scaling relation between the relaxation time and the correlation length $\tau\sim\xi^z$, and another one between the correlation length and the reduced temperature, $\xi\sim\tilde{T}^{-\nu}$, yield to the power-law \cite{Binder86}
\begin{equation}
\label{eq:RelaxTime}
\tau\sim\tilde{T}^{-z\nu}.
\end{equation}

Fitting the average relaxation time to Eq.~(\ref{eq:RelaxTime}) in the range $0.1\leq\tilde{T}\leq1$, we obtain the line in Fig.~\ref{fig:RelaxTime}. Following the same procedure for several $p$ values, we obtain the dynamic critical exponent $z\nu$ as a function of AFM bond concentration $p$, as presented in Fig.~\ref{fig:DynamicExp}. We observe that $z\nu$ increases with increasing $p$.

\begin{figure}[t!]
\includegraphics[scale=1.0]{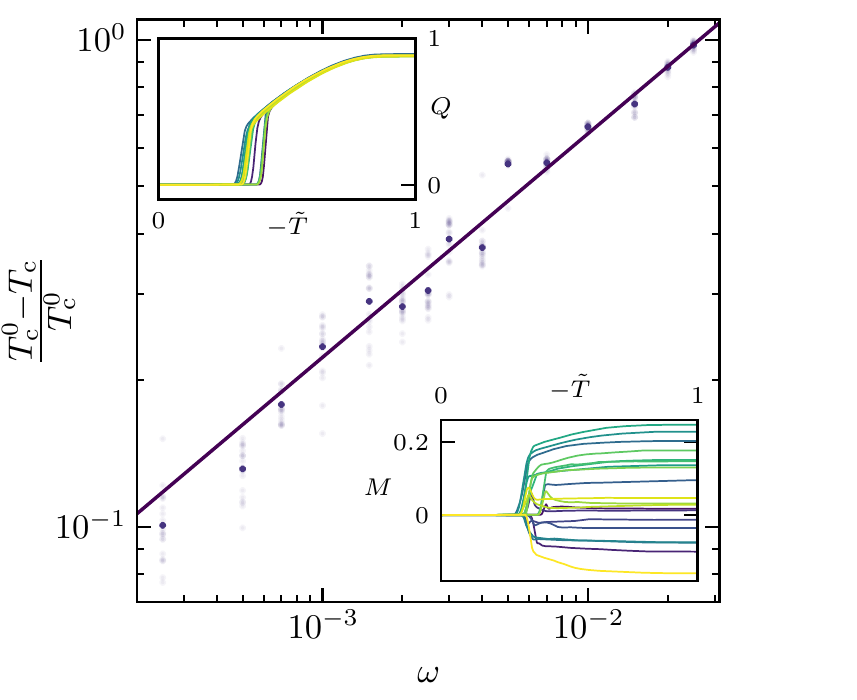}
\caption{\label{fig:CritTemp}Reduced difference $(T_\text{c}^0-T_\text{c})/T_\text{c}^0$ between the apparent dynamic ($T_\text{c}$) and the true non-dynamic ($T_\text{c}^0$) critical temperatures, as a function of cooling rate $\omega$, for $p=0.25$. Faded markers show the results for $20$ distinct realizations of AFM bond distributions, while the dark markers show averages over those random replicas. The line shows the best fit to the power-law Eq.~(\ref{eq:CritTemp}). The average magnetization, $M=L^{-3}\sum_im_i$ and the SG order parameter $Q=L^{-3}\sum_im_i^2$ are shown in the bottom-right and top-left insets as functions of $-\tilde{T}$ for all $20$ realizations for $\omega=0.0025$.}
\end{figure}

\section{Finite temperature sweep rate}

To determine the effect of finite cooling rate
\begin{equation}
\label{eq:SweepRate}
\omega=-\frac{dT}{dt}\neq0
\end{equation}
on the critical temperature $T_\text{c}(\omega)$, we start with a fully relaxed realization of the system at a fixed AFM bond distribution at a high temperature, $T=100$. The system is then cooled with a rate of change in temperature $-dT$ per time step. As defined above, one time step $dt$ corresponds to updating $L^3$ randomly chosen sites according to Eq.~(\ref{eq:Method}). We repeat the procedure for $\omega$ spanning two orders of magnitude in the range $[0.00025, 0.025]$, and for various $p$. For each set of values for $\omega$ and $p$, we generate $20$ distinct realizations of AFM bond distributions and work with the averages over those replicas.

For finite $\omega\neq0$, the system will not find enough time to fully relax to an ordered state when it reaches at the true non-dynamic critical temperature $T_\text{c}^0=T_\text{c}(\omega=0)$. And when it reaches an ordered state, the temperature will be already cooled below $T_\text{c}^0$. Thus, the apparent critical temperature $T_\text{c}(\omega)$ is expected to be lower than $T_\text{c}^0$, and to shift down with increasing cooling rate $\omega$.

The reduced difference $(T_\text{c}^0-T_\text{c})/T_\text{c}^0$ between the apparent dynamic $T_\text{c}$ and the true non-dynamic $T_\text{c}^0$ is plotted in Fig.~\ref{fig:CritTemp} in log-log axes as a function of $\omega$, for a representative value $p=0.25$, while we obtain the same behavior for other $p$ values. The average magnetization, $M=L^{-3}\sum_im_i$ (shown in the bottom-right inset for $20$ distinct realizations for $p=0.25$ and $\omega=0.0025$), is not suitable to determine $T_\text{c}$ for PM-SG phase transition at $p>p_\text{c}$. Instead, the SG order parameter $Q=L^{-3}\sum_im_i^2$ (shown in the top-left inset for $20$ distinct realizations for $p=0.25$ and $\omega=0.0025$) is used for the whole range of $p$, implementing the fact that $Q=0$ for $T>T_\text{c}$ and $Q>0$ for $T<T_\text{c}$. We checked that for $\omega=0$, this condition yields the same equilibrium PM phase boundary as obtained from hysteretic behavior (see Fig.~\ref{fig:PhaseDiag}).

\begin{figure}[t!]
\includegraphics[scale=1.0]{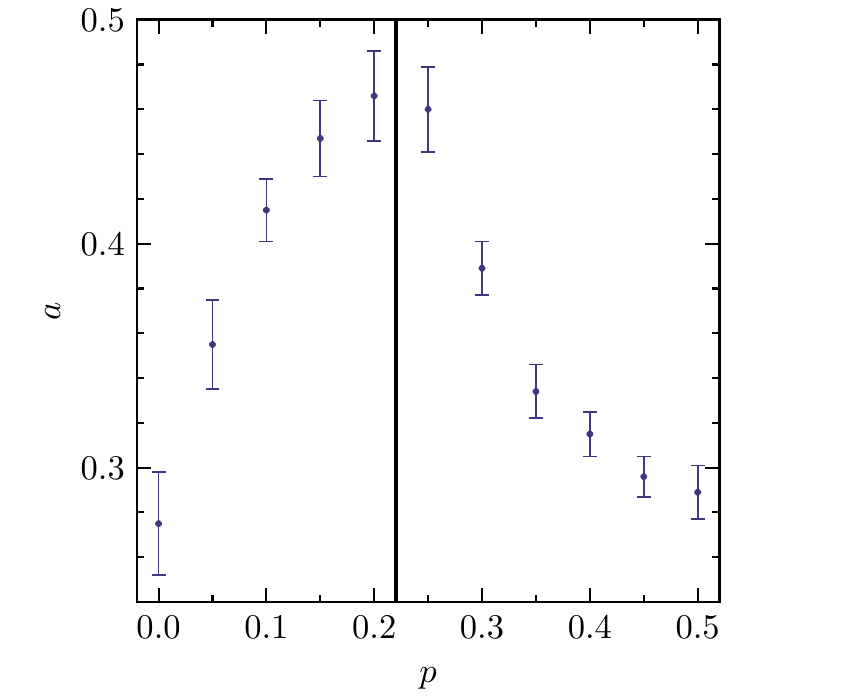}
\caption{\label{fig:SweepRateExp} Temperature sweep rate exponent $a$ of Eq.~(\ref{eq:CritTemp}) as a function of AFM bond fraction $p$. Error bars show the standard deviation of the fit procedure to Eq.~(\ref{eq:CritTemp}). Vertical line shows the critical $p_\text{c}\approx0.22$ between the ordered FM and SG phases  (see Fig.~\ref{fig:PhaseDiag}).}
\end{figure}

Particularly at higher $\omega$, the reduced difference in critical temperatures shows a clear power-law behavior,
\begin{equation}
\label{eq:CritTemp}
\frac{T_\text{c}^0-T_\text{c}}{T_\text{c}^0}\sim\omega^{a},
\end{equation}
as shown by the fit of average data to Eq.~(\ref{eq:CritTemp}), resulting in the line in Fig.~\ref{fig:CritTemp}. Such is the case for all other $p$ values, and the resulting exponents are shown in Fig.~\ref{fig:SweepRateExp} for various $p$. We clearly observe the peak in the exponent $a$ at the FM-SG phase transition at $p_\text{c}\approx0.22$.

\section{Conclusion}

We simulated the phase transition dynamics of $\pm J$ Ising spin glass model on a simple cubic lattice by the frustration preserving hard-spin mean-field theory. We obtained the dynamic exponents $z\nu$ for the relaxation time, and $a$ for the cooling rate dependence of critical temperature, $T_\textit{c}(\omega)$.

We found that the dynamic exponent $z\nu$ varies in the range of about $0.5$ to $0.7$, which is an order of magnitude smaller than the values reported by AC magnetometry experiments \cite{Svedlindh87a, Svedlindh87b, Mauger88, Gunnarsson88, Nam00, Quilliam08, Wang10, Perovic11, Perovic13, Scholz16} and stochastic simulations \cite{Ogielski85a, Ogielski85b}. Such is the pitfall of mean-field theories in general. Neglecting the long range correlations close to critical points where correlation length diverges, mean-field theories exhibit inadequate values for critical temperatures, amplitudes and exponents. However, they offer qualitative pictures of phase transitions. In this particular case, although our values for the dynamic exponent $z\nu$ do not agree with previous results, the qualitative feature that it increases with increasing AFM bond ratio $p$ is acceptable. This can be understood on the basis of increasing relaxation time with increasing frustration (increasing raggedness and number of local minima in the free energy landscape) with increasing $p$. This may also explain the various experimental values for $z\nu$ ranging in the range from $5.6\pm0.1$ for Sn$_{0.9}$Fe$_{3.1}$N \cite{Scholz16} to $10.4\pm1.0$ for Fe$_{0.5}$Mn$_{0.5}$TiO$_3$ \cite{Svedlindh87b}.

Similarly, although the values might be incorrect, the cooling rate exponent $a$ shows a peak at $p=p_\text{c}\approx0.22$, which clearly indicates the FM-SG transition. In these simulations, we initialized the cooling from a high temperature $T=100$. We observed that starting the cooling from $T=10$, changes our results only within the standard deviations even for the highest cooling rates, and hence, we infer that $T=100$ is a safe high-temperature starting point. Nonetheless, the effect of initial temperature on the dynamic properties, is another interesting direction to cover. We expect that as the starting temperature gets smaller, the apparent dynamic critical temperatures will decrease further. This is because the system will find even smaller times to relax to an ordered state, and by the time it relaxes, the temperature will be even more smaller than the true non-dynamic $T_\text{c}^0$.


\begin{acknowledgments}
I thank Dr.\,Aykut Erba\c{s} for valuable discussions and for careful reading of the manuscript. Numerical calculations were run by a machine partially supported by the 2232 TÜBİTAK Reintegration Grant of project \#115C135.
\end{acknowledgments}

\bibliography{pmJdynamics}

\begin{thebibliography}{36}%
\makeatletter
\providecommand \@ifxundefined [1]{%
 \@ifx{#1\undefined}
}%
\providecommand \@ifnum [1]{%
 \ifnum #1\expandafter \@firstoftwo
 \else \expandafter \@secondoftwo
 \fi
}%
\providecommand \@ifx [1]{%
 \ifx #1\expandafter \@firstoftwo
 \else \expandafter \@secondoftwo
 \fi
}%
\providecommand \natexlab [1]{#1}%
\providecommand \enquote  [1]{``#1''}%
\providecommand \bibnamefont  [1]{#1}%
\providecommand \bibfnamefont [1]{#1}%
\providecommand \citenamefont [1]{#1}%
\providecommand \href@noop [0]{\@secondoftwo}%
\providecommand \href [0]{\begingroup \@sanitize@url \@href}%
\providecommand \@href[1]{\@@startlink{#1}\@@href}%
\providecommand \@@href[1]{\endgroup#1\@@endlink}%
\providecommand \@sanitize@url [0]{\catcode `\\12\catcode `\$12\catcode
  `\&12\catcode `\#12\catcode `\^12\catcode `\_12\catcode `\%12\relax}%
\providecommand \@@startlink[1]{}%
\providecommand \@@endlink[0]{}%
\providecommand \url  [0]{\begingroup\@sanitize@url \@url }%
\providecommand \@url [1]{\endgroup\@href {#1}{\urlprefix }}%
\providecommand \urlprefix  [0]{URL }%
\providecommand \Eprint [0]{\href }%
\providecommand \doibase [0]{http://dx.doi.org/}%
\providecommand \selectlanguage [0]{\@gobble}%
\providecommand \bibinfo  [0]{\@secondoftwo}%
\providecommand \bibfield  [0]{\@secondoftwo}%
\providecommand \translation [1]{[#1]}%
\providecommand \BibitemOpen [0]{}%
\providecommand \bibitemStop [0]{}%
\providecommand \bibitemNoStop [0]{.\EOS\space}%
\providecommand \EOS [0]{\spacefactor3000\relax}%
\providecommand \BibitemShut  [1]{\csname bibitem#1\endcsname}%
\let\auto@bib@innerbib\@empty
\bibitem [{\citenamefont {Dupuis}\ \emph {et~al.}(2001)\citenamefont {Dupuis},
  \citenamefont {Vincent}, \citenamefont {Bouchaud}, \citenamefont {Hammann},
  \citenamefont {Ito},\ and\ \citenamefont {Katori}}]{Dupuis01}%
  \BibitemOpen
  \bibfield  {author} {\bibinfo {author} {\bibfnamefont {V.}~\bibnamefont
  {Dupuis}}, \bibinfo {author} {\bibfnamefont {E.}~\bibnamefont {Vincent}},
  \bibinfo {author} {\bibfnamefont {J.-P.}\ \bibnamefont {Bouchaud}}, \bibinfo
  {author} {\bibfnamefont {J.}~\bibnamefont {Hammann}}, \bibinfo {author}
  {\bibfnamefont {A.}~\bibnamefont {Ito}}, \ and\ \bibinfo {author}
  {\bibfnamefont {H.~A.}\ \bibnamefont {Katori}},\ }\href {\doibase
  10.1103/PhysRevB.64.174204} {\bibfield  {journal} {\bibinfo  {journal} {Phys.
  Rev. B}\ }\textbf {\bibinfo {volume} {64}},\ \bibinfo {pages} {174204}
  (\bibinfo {year} {2001})}\BibitemShut {NoStop}%
\bibitem [{\citenamefont {Perovic}\ \emph {et~al.}(2013)\citenamefont
  {Perovic}, \citenamefont {Kusigerski}, \citenamefont {Spasojevic},
  \citenamefont {Mrakovic}, \citenamefont {Blanusa}, \citenamefont {Zentkova},\
  and\ \citenamefont {Mihalik}}]{Perovic13}%
  \BibitemOpen
  \bibfield  {author} {\bibinfo {author} {\bibfnamefont {M.}~\bibnamefont
  {Perovic}}, \bibinfo {author} {\bibfnamefont {V.}~\bibnamefont {Kusigerski}},
  \bibinfo {author} {\bibfnamefont {V.}~\bibnamefont {Spasojevic}}, \bibinfo
  {author} {\bibfnamefont {A.}~\bibnamefont {Mrakovic}}, \bibinfo {author}
  {\bibfnamefont {J.}~\bibnamefont {Blanusa}}, \bibinfo {author} {\bibfnamefont
  {M.}~\bibnamefont {Zentkova}}, \ and\ \bibinfo {author} {\bibfnamefont
  {M.}~\bibnamefont {Mihalik}},\ }\href {\doibase
  10.1088/0022-3727/46/16/165001} {\bibfield  {journal} {\bibinfo  {journal}
  {J. Phys. D}\ }\textbf {\bibinfo {volume} {46}},\ \bibinfo {pages} {165001}
  (\bibinfo {year} {2013})}\BibitemShut {NoStop}%
\bibitem [{\citenamefont {Svedlindh}\ \emph
  {et~al.}(1987{\natexlab{a}})\citenamefont {Svedlindh}, \citenamefont
  {Gunnarsson}, \citenamefont {Nordblad}, \citenamefont {Lundgren},
  \citenamefont {Ito},\ and\ \citenamefont {Aruga}}]{Svedlindh87a}%
  \BibitemOpen
  \bibfield  {author} {\bibinfo {author} {\bibfnamefont {P.}~\bibnamefont
  {Svedlindh}}, \bibinfo {author} {\bibfnamefont {K.}~\bibnamefont
  {Gunnarsson}}, \bibinfo {author} {\bibfnamefont {P.}~\bibnamefont
  {Nordblad}}, \bibinfo {author} {\bibfnamefont {L.}~\bibnamefont {Lundgren}},
  \bibinfo {author} {\bibfnamefont {A.}~\bibnamefont {Ito}}, \ and\ \bibinfo
  {author} {\bibfnamefont {H.}~\bibnamefont {Aruga}},\ }\href {\doibase
  10.1016/0304-8853(87)90329-5} {\bibfield  {journal} {\bibinfo  {journal} {J.
  Magn. Magn. Mater.}\ }\textbf {\bibinfo {volume} {71}},\ \bibinfo {pages}
  {22} (\bibinfo {year} {1987}{\natexlab{a}})}\BibitemShut {NoStop}%
\bibitem [{\citenamefont {Toulouse}(1977)}]{Toulouse77}%
  \BibitemOpen
  \bibfield  {author} {\bibinfo {author} {\bibfnamefont {G.}~\bibnamefont
  {Toulouse}},\ }\href@noop {} {\bibfield  {journal} {\bibinfo  {journal}
  {Communications on Physics}\ }\textbf {\bibinfo {volume} {2}},\ \bibinfo
  {pages} {115} (\bibinfo {year} {1977})}\BibitemShut {NoStop}%
\bibitem [{\citenamefont {Banavar}\ \emph {et~al.}(1991)\citenamefont
  {Banavar}, \citenamefont {Cieplak},\ and\ \citenamefont
  {Maritan}}]{Banavar91}%
  \BibitemOpen
  \bibfield  {author} {\bibinfo {author} {\bibfnamefont {J.~R.}\ \bibnamefont
  {Banavar}}, \bibinfo {author} {\bibfnamefont {M.}~\bibnamefont {Cieplak}}, \
  and\ \bibinfo {author} {\bibfnamefont {A.}~\bibnamefont {Maritan}},\ }\href
  {\doibase 10.1103/PhysRevLett.67.1807} {\bibfield  {journal} {\bibinfo
  {journal} {Phys. Rev. Lett.}\ }\textbf {\bibinfo {volume} {67}},\ \bibinfo
  {pages} {1807} (\bibinfo {year} {1991})}\BibitemShut {NoStop}%
\bibitem [{\citenamefont {Netz}\ and\ \citenamefont
  {Berker}(1991{\natexlab{a}})}]{Netz91a}%
  \BibitemOpen
  \bibfield  {author} {\bibinfo {author} {\bibfnamefont {R.~R.}\ \bibnamefont
  {Netz}}\ and\ \bibinfo {author} {\bibfnamefont {A.~N.}\ \bibnamefont
  {Berker}},\ }\href {\doibase 10.1103/PhysRevLett.66.377} {\bibfield
  {journal} {\bibinfo  {journal} {Phys. Rev. Lett.}\ }\textbf {\bibinfo
  {volume} {66}},\ \bibinfo {pages} {377} (\bibinfo {year}
  {1991}{\natexlab{a}})}\BibitemShut {NoStop}%
\bibitem [{\citenamefont {Netz}\ and\ \citenamefont
  {Berker}(1991{\natexlab{b}})}]{Netz91b}%
  \BibitemOpen
  \bibfield  {author} {\bibinfo {author} {\bibfnamefont {R.~R.}\ \bibnamefont
  {Netz}}\ and\ \bibinfo {author} {\bibfnamefont {A.~N.}\ \bibnamefont
  {Berker}},\ }\href {\doibase 10.1103/PhysRevLett.67.1808} {\bibfield
  {journal} {\bibinfo  {journal} {Phys. Rev. Lett.}\ }\textbf {\bibinfo
  {volume} {67}},\ \bibinfo {pages} {1808} (\bibinfo {year}
  {1991}{\natexlab{b}})}\BibitemShut {NoStop}%
\bibitem [{\citenamefont {Netz}\ and\ \citenamefont
  {Berker}(1991{\natexlab{c}})}]{Netz91c}%
  \BibitemOpen
  \bibfield  {author} {\bibinfo {author} {\bibfnamefont {R.~R.}\ \bibnamefont
  {Netz}}\ and\ \bibinfo {author} {\bibfnamefont {A.~N.}\ \bibnamefont
  {Berker}},\ }\href {\doibase 10.1063/1.350050} {\bibfield  {journal}
  {\bibinfo  {journal} {J. Appl. Phys.}\ }\textbf {\bibinfo {volume} {70}},\
  \bibinfo {pages} {6074} (\bibinfo {year} {1991}{\natexlab{c}})}\BibitemShut
  {NoStop}%
\bibitem [{\citenamefont {Netz}(1992)}]{Netz92}%
  \BibitemOpen
  \bibfield  {author} {\bibinfo {author} {\bibfnamefont {R.}~\bibnamefont
  {Netz}},\ }\href {\doibase 10.1103/PhysRevB.46.1209} {\bibfield  {journal}
  {\bibinfo  {journal} {Phys. Rev. B}\ }\textbf {\bibinfo {volume} {46}},\
  \bibinfo {pages} {1209} (\bibinfo {year} {1992})}\BibitemShut {NoStop}%
\bibitem [{\citenamefont {Netz}(1993)}]{Netz93}%
  \BibitemOpen
  \bibfield  {author} {\bibinfo {author} {\bibfnamefont {R.}~\bibnamefont
  {Netz}},\ }\href {\doibase 10.1103/PhysRevB.48.16113} {\bibfield  {journal}
  {\bibinfo  {journal} {Phys. Rev. B}\ }\textbf {\bibinfo {volume} {48}},\
  \bibinfo {pages} {16113} (\bibinfo {year} {1993})}\BibitemShut {NoStop}%
\bibitem [{\citenamefont {Ames}\ and\ \citenamefont {McKay}(1994)}]{Ames94}%
  \BibitemOpen
  \bibfield  {author} {\bibinfo {author} {\bibfnamefont {E.~A.}\ \bibnamefont
  {Ames}}\ and\ \bibinfo {author} {\bibfnamefont {S.~R.}\ \bibnamefont
  {McKay}},\ }\href {\doibase 10.1063/1.358350} {\bibfield  {journal} {\bibinfo
   {journal} {J. Appl. Phys.}\ }\textbf {\bibinfo {volume} {76}},\ \bibinfo
  {pages} {6197} (\bibinfo {year} {1994})}\BibitemShut {NoStop}%
\bibitem [{\citenamefont {Berker}\ \emph {et~al.}(1994)\citenamefont {Berker},
  \citenamefont {Kabak\c{c}{\i}o\u{g}lu}, \citenamefont {Netz},\ and\
  \citenamefont {Yalab{\i}k}}]{Berker94}%
  \BibitemOpen
  \bibfield  {author} {\bibinfo {author} {\bibfnamefont {A.}~\bibnamefont
  {Berker}}, \bibinfo {author} {\bibfnamefont {A.}~\bibnamefont
  {Kabak\c{c}{\i}o\u{g}lu}}, \bibinfo {author} {\bibfnamefont {R.}~\bibnamefont
  {Netz}}, \ and\ \bibinfo {author} {\bibfnamefont {M.}~\bibnamefont
  {Yalab{\i}k}},\ }\href@noop {} {\bibfield  {journal} {\bibinfo  {journal}
  {Turk. J. Phys.}\ }\textbf {\bibinfo {volume} {18}},\ \bibinfo {pages} {354}
  (\bibinfo {year} {1994})}\BibitemShut {NoStop}%
\bibitem [{\citenamefont {Kabak\c{c}{\i}o\u{g}lu}\ \emph
  {et~al.}(1994)\citenamefont {Kabak\c{c}{\i}o\u{g}lu}, \citenamefont
  {Berker},\ and\ \citenamefont {Yalab{\i}k}}]{Kabakcioglu94}%
  \BibitemOpen
  \bibfield  {author} {\bibinfo {author} {\bibfnamefont {A.}~\bibnamefont
  {Kabak\c{c}{\i}o\u{g}lu}}, \bibinfo {author} {\bibfnamefont {A.~N.}\
  \bibnamefont {Berker}}, \ and\ \bibinfo {author} {\bibfnamefont {M.~C.}\
  \bibnamefont {Yalab{\i}k}},\ }\href {\doibase 10.1103/PhysRevE.49.2680}
  {\bibfield  {journal} {\bibinfo  {journal} {Phys. Rev. E}\ }\textbf {\bibinfo
  {volume} {49}},\ \bibinfo {pages} {2680} (\bibinfo {year}
  {1994})}\BibitemShut {NoStop}%
\bibitem [{\citenamefont {Akg{\"{u}}\c{c}}\ and\ \citenamefont
  {Yalab{\i}k}(1995)}]{Akguc95}%
  \BibitemOpen
  \bibfield  {author} {\bibinfo {author} {\bibfnamefont {G.~B.}\ \bibnamefont
  {Akg{\"{u}}\c{c}}}\ and\ \bibinfo {author} {\bibfnamefont {M.~C.}\
  \bibnamefont {Yalab{\i}k}},\ }\href {\doibase 10.1103/PhysRevE.51.2636}
  {\bibfield  {journal} {\bibinfo  {journal} {Phys. Rev. E}\ }\textbf {\bibinfo
  {volume} {51}},\ \bibinfo {pages} {2636} (\bibinfo {year}
  {1995})}\BibitemShut {NoStop}%
\bibitem [{\citenamefont {Tesiero}\ and\ \citenamefont
  {McKay}(1996)}]{Tesiero96}%
  \BibitemOpen
  \bibfield  {author} {\bibinfo {author} {\bibfnamefont {J.~E.}\ \bibnamefont
  {Tesiero}}\ and\ \bibinfo {author} {\bibfnamefont {S.~R.}\ \bibnamefont
  {McKay}},\ }\href {\doibase 10.1063/1.362052} {\bibfield  {journal} {\bibinfo
   {journal} {J. Appl. Phys.}\ }\textbf {\bibinfo {volume} {79}},\ \bibinfo
  {pages} {6146} (\bibinfo {year} {1996})}\BibitemShut {NoStop}%
\bibitem [{\citenamefont {Monroe}(1997)}]{Monroe97}%
  \BibitemOpen
  \bibfield  {author} {\bibinfo {author} {\bibfnamefont {J.~L.}\ \bibnamefont
  {Monroe}},\ }\href {\doibase 10.1016/S0375-9601(97)00188-6} {\bibfield
  {journal} {\bibinfo  {journal} {Phys. Lett. A}\ }\textbf {\bibinfo {volume}
  {230}},\ \bibinfo {pages} {111} (\bibinfo {year} {1997})}\BibitemShut
  {NoStop}%
\bibitem [{\citenamefont {Pelizzola}\ and\ \citenamefont
  {Pretti}(1999)}]{Pelizzola99}%
  \BibitemOpen
  \bibfield  {author} {\bibinfo {author} {\bibfnamefont {A.}~\bibnamefont
  {Pelizzola}}\ and\ \bibinfo {author} {\bibfnamefont {M.}~\bibnamefont
  {Pretti}},\ }\href {\doibase 10.1103/PhysRevB.60.10134} {\bibfield  {journal}
  {\bibinfo  {journal} {Phys. Rev. B}\ }\textbf {\bibinfo {volume} {60}},\
  \bibinfo {pages} {10134} (\bibinfo {year} {1999})}\BibitemShut {NoStop}%
\bibitem [{\citenamefont {Kabak\c{c}{\i}o\u{g}lu}(2000)}]{Kabakcioglu00}%
  \BibitemOpen
  \bibfield  {author} {\bibinfo {author} {\bibfnamefont {A.}~\bibnamefont
  {Kabak\c{c}{\i}o\u{g}lu}},\ }\href {\doibase 10.1103/PhysRevE.61.3366}
  {\bibfield  {journal} {\bibinfo  {journal} {Phys. Rev. E}\ }\textbf {\bibinfo
  {volume} {61}},\ \bibinfo {pages} {3366} (\bibinfo {year}
  {2000})}\BibitemShut {NoStop}%
\bibitem [{\citenamefont {Kaya}\ and\ \citenamefont {Berker}(2000)}]{Kaya00}%
  \BibitemOpen
  \bibfield  {author} {\bibinfo {author} {\bibfnamefont {H.}~\bibnamefont
  {Kaya}}\ and\ \bibinfo {author} {\bibfnamefont {A.~N.}\ \bibnamefont
  {Berker}},\ }\href {\doibase 10.1103/PhysRevE.62.R1469} {\bibfield  {journal}
  {\bibinfo  {journal} {Phys. Rev. E}\ }\textbf {\bibinfo {volume} {62}},\
  \bibinfo {pages} {R1469} (\bibinfo {year} {2000})}\BibitemShut {NoStop}%
\bibitem [{\citenamefont {Y{\"{u}}cesoy}\ and\ \citenamefont
  {Berker}(2007)}]{Yucesoy07}%
  \BibitemOpen
  \bibfield  {author} {\bibinfo {author} {\bibfnamefont {B.}~\bibnamefont
  {Y{\"{u}}cesoy}}\ and\ \bibinfo {author} {\bibfnamefont {A.~N.}\ \bibnamefont
  {Berker}},\ }\href {\doibase 10.1103/PhysRevB.76.014417} {\bibfield
  {journal} {\bibinfo  {journal} {Phys. Rev. B}\ }\textbf {\bibinfo {volume}
  {76}},\ \bibinfo {pages} {014417} (\bibinfo {year} {2007})}\BibitemShut
  {NoStop}%
\bibitem [{\citenamefont {Robinson}\ \emph {et~al.}(2011)\citenamefont
  {Robinson}, \citenamefont {Feldman},\ and\ \citenamefont
  {McKay}}]{Robinson11}%
  \BibitemOpen
  \bibfield  {author} {\bibinfo {author} {\bibfnamefont {M.~D.}\ \bibnamefont
  {Robinson}}, \bibinfo {author} {\bibfnamefont {D.~P.}\ \bibnamefont
  {Feldman}}, \ and\ \bibinfo {author} {\bibfnamefont {S.~R.}\ \bibnamefont
  {McKay}},\ }\href {\doibase 10.1063/1.3608120} {\bibfield  {journal}
  {\bibinfo  {journal} {Chaos}\ }\textbf {\bibinfo {volume} {21}},\ \bibinfo
  {pages} {037114} (\bibinfo {year} {2011})}\BibitemShut {NoStop}%
\bibitem [{\citenamefont {\c{C}a\u{g}lar}\ and\ \citenamefont
  {Berker}(2011)}]{Caglar11}%
  \BibitemOpen
  \bibfield  {author} {\bibinfo {author} {\bibfnamefont {T.}~\bibnamefont
  {\c{C}a\u{g}lar}}\ and\ \bibinfo {author} {\bibfnamefont {A.~N.}\
  \bibnamefont {Berker}},\ }\href {\doibase 10.1103/PhysRevE.84.051129}
  {\bibfield  {journal} {\bibinfo  {journal} {Phys. Rev. E}\ }\textbf {\bibinfo
  {volume} {84}},\ \bibinfo {pages} {051129} (\bibinfo {year}
  {2011})}\BibitemShut {NoStop}%
\bibitem [{\citenamefont {\c{C}a\u{g}lar}\ and\ \citenamefont
  {Berker}(2015)}]{Caglar15}%
  \BibitemOpen
  \bibfield  {author} {\bibinfo {author} {\bibfnamefont {T.}~\bibnamefont
  {\c{C}a\u{g}lar}}\ and\ \bibinfo {author} {\bibfnamefont {A.~N.}\
  \bibnamefont {Berker}},\ }\href {\doibase 10.1103/PhysRevE.92.062131}
  {\bibfield  {journal} {\bibinfo  {journal} {Phys. Rev. E}\ }\textbf {\bibinfo
  {volume} {92}},\ \bibinfo {pages} {062131} (\bibinfo {year}
  {2015})}\BibitemShut {NoStop}%
\bibitem [{\citenamefont {Sar{\i}yer}\ \emph {et~al.}(2012)\citenamefont
  {Sar{\i}yer}, \citenamefont {Kabak\c{c}{\i}o\u{g}lu},\ and\ \citenamefont
  {Berker}}]{Sariyer12}%
  \BibitemOpen
  \bibfield  {author} {\bibinfo {author} {\bibfnamefont {O.~S.}\ \bibnamefont
  {Sar{\i}yer}}, \bibinfo {author} {\bibfnamefont {A.}~\bibnamefont
  {Kabak\c{c}{\i}o\u{g}lu}}, \ and\ \bibinfo {author} {\bibfnamefont {A.~N.}\
  \bibnamefont {Berker}},\ }\href {\doibase 10.1103/PhysRevE.86.041107}
  {\bibfield  {journal} {\bibinfo  {journal} {Phys. Rev. E}\ }\textbf {\bibinfo
  {volume} {86}},\ \bibinfo {pages} {041107} (\bibinfo {year}
  {2012})}\BibitemShut {NoStop}%
\bibitem [{\citenamefont {Ozeki}\ and\ \citenamefont
  {Nishimori}(1987)}]{Ozeki87}%
  \BibitemOpen
  \bibfield  {author} {\bibinfo {author} {\bibfnamefont {Y.}~\bibnamefont
  {Ozeki}}\ and\ \bibinfo {author} {\bibfnamefont {H.}~\bibnamefont
  {Nishimori}},\ }\href {\doibase 10.1143/JPSJ.56.1568} {\bibfield  {journal}
  {\bibinfo  {journal} {J. Phys. Soc. Jpn.}\ }\textbf {\bibinfo {volume}
  {56}},\ \bibinfo {pages} {1568} (\bibinfo {year} {1987})}\BibitemShut
  {NoStop}%
\bibitem [{\citenamefont {Binder}\ and\ \citenamefont
  {Young}(1986)}]{Binder86}%
  \BibitemOpen
  \bibfield  {author} {\bibinfo {author} {\bibfnamefont {K.}~\bibnamefont
  {Binder}}\ and\ \bibinfo {author} {\bibfnamefont {A.~P.}\ \bibnamefont
  {Young}},\ }\href {\doibase 10.1103/RevModPhys.58.801} {\bibfield  {journal}
  {\bibinfo  {journal} {Rev. Mod. Phys.}\ }\textbf {\bibinfo {volume} {58}},\
  \bibinfo {pages} {801} (\bibinfo {year} {1986})}\BibitemShut {NoStop}%
\bibitem [{\citenamefont {Svedlindh}\ \emph
  {et~al.}(1987{\natexlab{b}})\citenamefont {Svedlindh}, \citenamefont
  {Granberg}, \citenamefont {Nordblad}, \citenamefont {Lundgren},\ and\
  \citenamefont {Chen}}]{Svedlindh87b}%
  \BibitemOpen
  \bibfield  {author} {\bibinfo {author} {\bibfnamefont {P.}~\bibnamefont
  {Svedlindh}}, \bibinfo {author} {\bibfnamefont {P.}~\bibnamefont {Granberg}},
  \bibinfo {author} {\bibfnamefont {P.}~\bibnamefont {Nordblad}}, \bibinfo
  {author} {\bibfnamefont {L.}~\bibnamefont {Lundgren}}, \ and\ \bibinfo
  {author} {\bibfnamefont {H.~S.}\ \bibnamefont {Chen}},\ }\href {\doibase
  10.1103/PhysRevB.35.268} {\bibfield  {journal} {\bibinfo  {journal} {Phys.
  Rev. B}\ }\textbf {\bibinfo {volume} {35}},\ \bibinfo {pages} {268} (\bibinfo
  {year} {1987}{\natexlab{b}})}\BibitemShut {NoStop}%
\bibitem [{\citenamefont {Mauger}\ \emph {et~al.}(1988)\citenamefont {Mauger},
  \citenamefont {Ferr\'e}, \citenamefont {Ayadi},\ and\ \citenamefont
  {Nordblad}}]{Mauger88}%
  \BibitemOpen
  \bibfield  {author} {\bibinfo {author} {\bibfnamefont {A.}~\bibnamefont
  {Mauger}}, \bibinfo {author} {\bibfnamefont {J.}~\bibnamefont {Ferr\'e}},
  \bibinfo {author} {\bibfnamefont {M.}~\bibnamefont {Ayadi}}, \ and\ \bibinfo
  {author} {\bibfnamefont {P.}~\bibnamefont {Nordblad}},\ }\href {\doibase
  10.1103/PhysRevB.37.9022} {\bibfield  {journal} {\bibinfo  {journal} {Phys.
  Rev. B}\ }\textbf {\bibinfo {volume} {37}},\ \bibinfo {pages} {9002}
  (\bibinfo {year} {1988})}\BibitemShut {NoStop}%
\bibitem [{\citenamefont {Gunnarsson}\ \emph {et~al.}(1988)\citenamefont
  {Gunnarsson}, \citenamefont {Svedlindh}, \citenamefont {Nordblad},
  \citenamefont {Lundgren}, \citenamefont {Aruga},\ and\ \citenamefont
  {Ito}}]{Gunnarsson88}%
  \BibitemOpen
  \bibfield  {author} {\bibinfo {author} {\bibfnamefont {K.}~\bibnamefont
  {Gunnarsson}}, \bibinfo {author} {\bibfnamefont {P.}~\bibnamefont
  {Svedlindh}}, \bibinfo {author} {\bibfnamefont {P.}~\bibnamefont {Nordblad}},
  \bibinfo {author} {\bibfnamefont {L.}~\bibnamefont {Lundgren}}, \bibinfo
  {author} {\bibfnamefont {H.}~\bibnamefont {Aruga}}, \ and\ \bibinfo {author}
  {\bibfnamefont {A.}~\bibnamefont {Ito}},\ }\href {\doibase
  10.1103/PhysRevLett.61.754} {\bibfield  {journal} {\bibinfo  {journal} {Phys.
  Rev. Lett.}\ }\textbf {\bibinfo {volume} {61}},\ \bibinfo {pages} {754}
  (\bibinfo {year} {1988})}\BibitemShut {NoStop}%
\bibitem [{\citenamefont {Nam}\ \emph {et~al.}(2000)\citenamefont {Nam},
  \citenamefont {Mathieu}, \citenamefont {Nordblad}, \citenamefont {Khiem},\
  and\ \citenamefont {Phuc}}]{Nam00}%
  \BibitemOpen
  \bibfield  {author} {\bibinfo {author} {\bibfnamefont {D.~N.~H.}\
  \bibnamefont {Nam}}, \bibinfo {author} {\bibfnamefont {R.}~\bibnamefont
  {Mathieu}}, \bibinfo {author} {\bibfnamefont {P.}~\bibnamefont {Nordblad}},
  \bibinfo {author} {\bibfnamefont {N.~V.}\ \bibnamefont {Khiem}}, \ and\
  \bibinfo {author} {\bibfnamefont {N.~X.}\ \bibnamefont {Phuc}},\ }\href
  {\doibase 10.1103/PhysRevB.62.8989} {\bibfield  {journal} {\bibinfo
  {journal} {Phys. Rev. B}\ }\textbf {\bibinfo {volume} {62}},\ \bibinfo
  {pages} {8989} (\bibinfo {year} {2000})}\BibitemShut {NoStop}%
\bibitem [{\citenamefont {Quilliam}\ \emph {et~al.}(2008)\citenamefont
  {Quilliam}, \citenamefont {Meng}, \citenamefont {Mugford},\ and\
  \citenamefont {Kycia}}]{Quilliam08}%
  \BibitemOpen
  \bibfield  {author} {\bibinfo {author} {\bibfnamefont {J.~A.}\ \bibnamefont
  {Quilliam}}, \bibinfo {author} {\bibfnamefont {S.}~\bibnamefont {Meng}},
  \bibinfo {author} {\bibfnamefont {C.~G.~A.}\ \bibnamefont {Mugford}}, \ and\
  \bibinfo {author} {\bibfnamefont {J.~B.}\ \bibnamefont {Kycia}},\ }\href
  {\doibase 10.1103/PhysRevLett.101.187204} {\bibfield  {journal} {\bibinfo
  {journal} {Phys. Rev. Lett.}\ }\textbf {\bibinfo {volume} {101}},\ \bibinfo
  {pages} {187204} (\bibinfo {year} {2008})}\BibitemShut {NoStop}%
\bibitem [{\citenamefont {Wang}\ \emph {et~al.}(2010)\citenamefont {Wang},
  \citenamefont {Tong}, \citenamefont {Sun}, \citenamefont {Zhu}, \citenamefont
  {Yang}, \citenamefont {Song},\ and\ \citenamefont {Dai}}]{Wang10}%
  \BibitemOpen
  \bibfield  {author} {\bibinfo {author} {\bibfnamefont {B.~S.}\ \bibnamefont
  {Wang}}, \bibinfo {author} {\bibfnamefont {P.}~\bibnamefont {Tong}}, \bibinfo
  {author} {\bibfnamefont {Y.~P.}\ \bibnamefont {Sun}}, \bibinfo {author}
  {\bibfnamefont {X.~B.}\ \bibnamefont {Zhu}}, \bibinfo {author} {\bibfnamefont
  {Z.~R.}\ \bibnamefont {Yang}}, \bibinfo {author} {\bibfnamefont {W.~H.}\
  \bibnamefont {Song}}, \ and\ \bibinfo {author} {\bibfnamefont {J.~M.}\
  \bibnamefont {Dai}},\ }\href {\doibase 10.1063/1.3469924} {\bibfield
  {journal} {\bibinfo  {journal} {Appl. Phys. Lett.}\ }\textbf {\bibinfo
  {volume} {97}},\ \bibinfo {pages} {042508} (\bibinfo {year}
  {2010})}\BibitemShut {NoStop}%
\bibitem [{\citenamefont {Perovic}\ \emph {et~al.}(2011)\citenamefont
  {Perovic}, \citenamefont {Mrakovic}, \citenamefont {Kusigerski},
  \citenamefont {Blanusa},\ and\ \citenamefont {Spasojevic}}]{Perovic11}%
  \BibitemOpen
  \bibfield  {author} {\bibinfo {author} {\bibfnamefont {M.}~\bibnamefont
  {Perovic}}, \bibinfo {author} {\bibfnamefont {A.}~\bibnamefont {Mrakovic}},
  \bibinfo {author} {\bibfnamefont {V.}~\bibnamefont {Kusigerski}}, \bibinfo
  {author} {\bibfnamefont {J.}~\bibnamefont {Blanusa}}, \ and\ \bibinfo
  {author} {\bibfnamefont {V.}~\bibnamefont {Spasojevic}},\ }\href {\doibase
  10.1007/s11051-011-0588-4} {\bibfield  {journal} {\bibinfo  {journal} {J.
  Nanopart. Res.}\ }\textbf {\bibinfo {volume} {13}},\ \bibinfo {pages} {6805}
  (\bibinfo {year} {2011})}\BibitemShut {NoStop}%
\bibitem [{\citenamefont {Scholz}\ and\ \citenamefont
  {Dronskowski}(2016)}]{Scholz16}%
  \BibitemOpen
  \bibfield  {author} {\bibinfo {author} {\bibfnamefont {T.}~\bibnamefont
  {Scholz}}\ and\ \bibinfo {author} {\bibfnamefont {R.}~\bibnamefont
  {Dronskowski}},\ }\href {\doibase 10.1063/1.4948984} {\bibfield  {journal}
  {\bibinfo  {journal} {AIP Adv.}\ }\textbf {\bibinfo {volume} {6}},\ \bibinfo
  {pages} {055107} (\bibinfo {year} {2016})}\BibitemShut {NoStop}%
\bibitem [{\citenamefont {Ogielski}\ and\ \citenamefont
  {Morgenstern}(1985)}]{Ogielski85a}%
  \BibitemOpen
  \bibfield  {author} {\bibinfo {author} {\bibfnamefont {A.~T.}\ \bibnamefont
  {Ogielski}}\ and\ \bibinfo {author} {\bibfnamefont {I.}~\bibnamefont
  {Morgenstern}},\ }\href {\doibase 10.1103/PhysRevLett.54.928} {\bibfield
  {journal} {\bibinfo  {journal} {Phys. Rev. Lett.}\ }\textbf {\bibinfo
  {volume} {54}},\ \bibinfo {pages} {928} (\bibinfo {year} {1985})}\BibitemShut
  {NoStop}%
\bibitem [{\citenamefont {Ogielski}(1985)}]{Ogielski85b}%
  \BibitemOpen
  \bibfield  {author} {\bibinfo {author} {\bibfnamefont {A.~T.}\ \bibnamefont
  {Ogielski}},\ }\href {\doibase 10.1103/PhysRevB.32.7384} {\bibfield
  {journal} {\bibinfo  {journal} {Phys. Rev. B}\ }\textbf {\bibinfo {volume}
  {32}},\ \bibinfo {pages} {7384} (\bibinfo {year} {1985})}\BibitemShut
  {NoStop}%
\end{thebibliography}%

\end{document}